\documentclass[twocolumn,english,aps,prb,twocolum,superscriptaddress,bibnotes,amsmath,amssymb,floatfix]{revtex4-1}
\usepackage[colorlinks=true,citecolor=blue,linkcolor=blue]{hyperref}
\usepackage{xcolor}
\usepackage{soul}
\usepackage[utf8]{inputenc}
\usepackage[english]{babel}
\usepackage{amsmath,amsfonts,amssymb}
\usepackage[T1]{fontenc}
\usepackage{url}

\usepackage{amsmath}
\usepackage{amsfonts}
\usepackage{amssymb}
\usepackage{graphicx}

\begin{document}
\title{Soliton generation in AlGaAs microresonators at room temperature}

\author{Lue Wu}
\thanks{These authors contributed equally to this work.}
\affiliation{T. J. Watson Laboratory of Applied Physics, California Institute of Technology, Pasadena, CA 91125, USA}

\author{Weiqiang Xie}
\thanks{These authors contributed equally to this work.}

\affiliation{Department of Electrical and Computer Engineering, University of California, Santa Barbara, Santa Barbara, California 93106, USA}

\author{Chao Xiang}
\affiliation{Department of Electrical and Computer Engineering, University of California, Santa Barbara, Santa Barbara, California 93106, USA}

\author{Lin Chang}
\affiliation{Department of Electrical and Computer Engineering, University of California, Santa Barbara, Santa Barbara, California 93106, USA}

\author{Yan Yu}
\affiliation{T. J. Watson Laboratory of Applied Physics, California Institute of Technology, Pasadena, CA 91125, USA}

\author{Hao-Jing Chen}
\affiliation{T. J. Watson Laboratory of Applied Physics, California Institute of Technology, Pasadena, CA 91125, USA}

\author{Yoshihisa Yamamoto}
\affiliation{Physics \& Informatics Laboratories, NTT Research, Inc., Sunnyvale, California 94085, USA}

\author{John E. Bowers}
\email[]{bowers@ece.ucsb.edu}
\affiliation{Department of Electrical and Computer Engineering, University of California, Santa Barbara, Santa Barbara, California 93106, USA}

\author{Kerry J. Vahala}
\email[]{vahala@caltech.edu}
\affiliation{T. J. Watson Laboratory of Applied Physics, California Institute of Technology, Pasadena, CA 91125, USA}

\author{Myoung-Gyun Suh}
\email[]{myoung-gyun.suh@ntt-research.com}
\affiliation{Physics \& Informatics Laboratories, NTT Research, Inc., Sunnyvale, California 94085, USA}

\maketitle

\noindent\textbf{Abstract}

\noindent  {\bf Chip-integrated optical frequency combs are attractive optical sources in comb applications requiring high-repetition-rate, low power consumption, or compact size. Spontaneous soliton formation via Kerr parametric oscillation is a promising generation principle in these frequency combs, and has been demonstrated in several material platforms over the past decade. Of these materials, AlGaAs has one of the largest Kerr nonlinearity coefficients allowing low pump threshold comb generation. However, bright soliton generation using this material has only been possible at cryogenic temperature because of the large thermo-optic effect at room temperature, which hinders stable access to the soliton regime. Here, we report self-stabilized single soliton generation in AlGaAs microresonators at room temperature by utilizing a rising soliton step in large free-spectral-range resonators. With sub-milliWatt optical pump power, 1 THz repetition-rate soliton generation is demonstrated. Perfect soliton crystal formation and soliton breather states are also observed. Besides the advantages of large optical nonlinearity, the devices are natural candidates for integration with III-V pump lasers. }

\medskip

\noindent\textbf{Introduction}

\noindent  
Optical frequency combs have revolutionized precision time and frequency metrology, and they find a wide range of applications in areas as diverse as spectroscopy, optical communications, distance measurement, and low-noise microwave generation \cite{DiddamsVahalaUdemCombReview}. Recent advances in chip-integration of optical frequency combs \cite{gaeta2019photonic,zhang2019broadband,Chao_Science2021,Bowers2022OFCreview} could accelerate widespread use of frequency combs outside of the laboratory environment. Taking advantage of chip-based high Q-factor microresonators  \cite{VahalaReview}, dissipative Kerr soliton (DKS) formation using the optical $\chi^{(3)}$ nonlinearity is a promising method of comb formation \cite{TJKDKSreview}. These soliton microcombs typically feature femtosecond pulse widths and repetition rates ranging from GHz to THz. They are accordingly useful for applications requiring high-repetition-rate, low power consumption, or small form factor. Because the $\chi^{(3)}$ nonlinearity exists in all dielectric materials, DKSs have been demonstrated in many materials including silicon nitride\cite{brasch2016photonic}, silica\cite{VahalaSilicaSoliton} , aluminum nitride\cite{liu2021aluminum}, and lithium niobate\cite{he2019self,gong2020near}.

Among the various materials, AlGaAs offers a combination of large $\chi^{(3)}$ nonlinearity and high refractive index for achieving $\mu$W-level Kerr parametric oscillation threshold \cite{moss2013new,DTU2016AlGaAs,LinAlGaAs_NC2020,Xie2020AlGaAs,AlGaAsSihetero_PR2022}. Moreoever, because of its compatibility with III-V semiconductor lasers, it has the potential to be integrated with pump lasers. However, the large thermo-optical effect of AlGaAs has prohibited soliton generation at room temperature, and either cooling to cryogenic temperature \cite{KartikAlGaAsLTDKS_LPR20} or using generation of dark soliton pulses \cite{shu2022microcomb,PKUAlGaAsDark_arXiv21} has been necessary to achieve stable microcombs. 

Here, we demonstrate stable room-temperature bright-soliton generation in AlGaAs microrosonators for the first time. Single soliton states are generated at repetition rates of 1 THz. The coherence of the resulting soliton microcombs is confirmed by beatnote measurements with a self-referenced fiber comb system, observation of perfect soliton crystals (PSC) with 3 THz repetition rate \cite{TJKPSC_NPhy19}, and frequency response measurements \cite{TJKSwitching_NPhy17}. Breather solitons  \cite{CYBreatherPRL2016,ErwanBreater_NC2017,MJBreather_NC2017} are also observed. 

\begin{figure*}[t!]
\centering
\includegraphics[width =0.85 \linewidth]{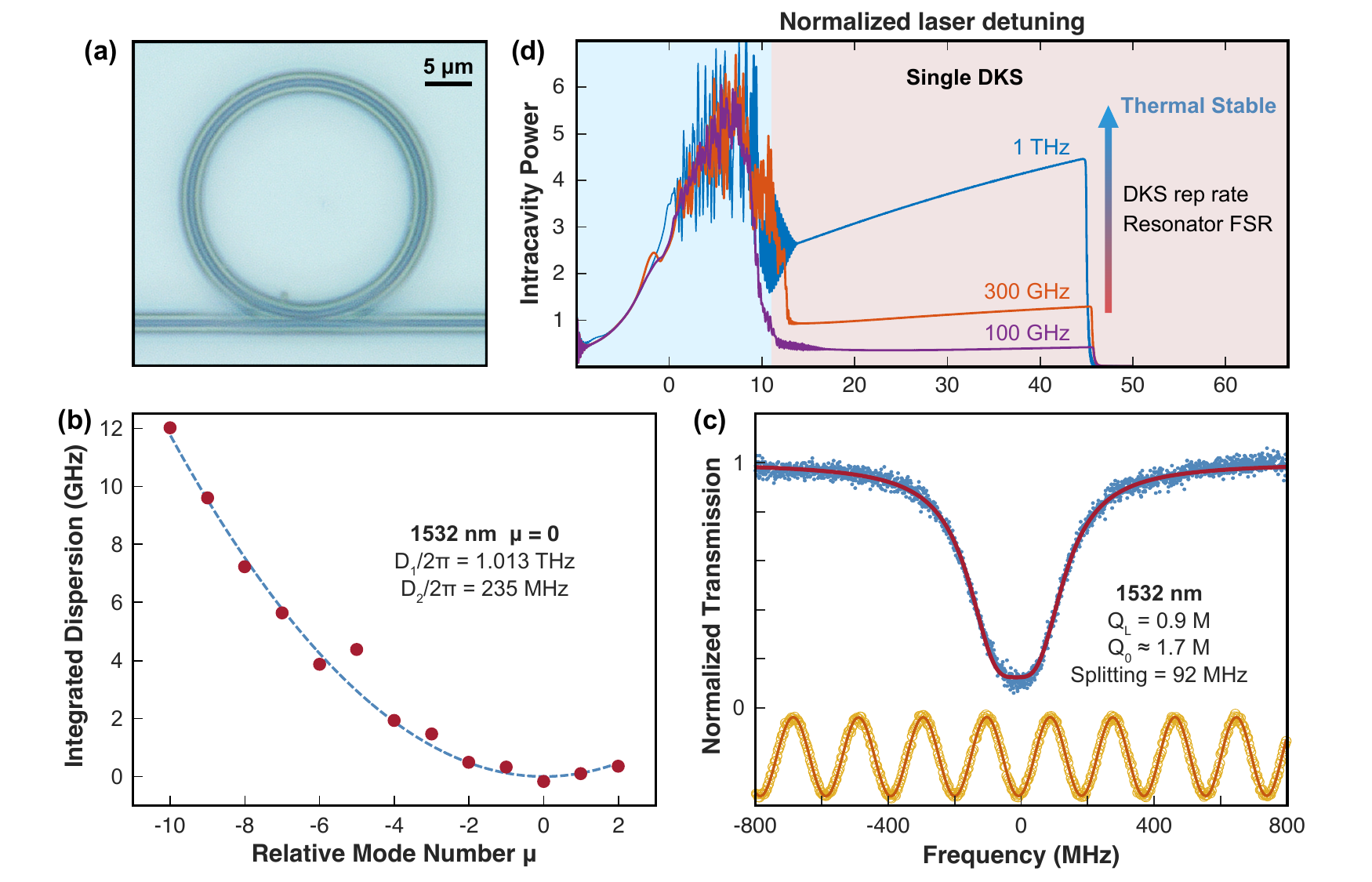}
\caption{1 THz FSR AlGaAs microresonator characterization. 
{\bf(a)}   Optical micrograph of device with radius 12.46 $\mu$m. 
{\bf(b)} Measured integrated frequency dispersion (red points) is plotted versus the relative mode number, $\mu$. To construct this plot, the center wavelength of each split mode is measured using an OSA and converted to frequency. The mode frequency is given by $\omega_{\mu}=\omega_0+\mu D_1+\frac{1}{2} D_2 \mu^2$, and the blue dashed curve is a fit using $D_1/2\pi =$ 1.0126 THz and $D_2/2\pi =$ 235 MHz. The measured modes span wavelengths from 1516 to 1620 nm, and $\mu=0$ corresponds to the pump mode wavelength at 1532 nm. A slight avoided mode crossing near $\mu=-5$ originates from TE/TM  mode hybridization.
{\bf(c)} Resonance linewidth measurement of the pumped mode at 1532 nm. Upper trace is the transmission spectrum (blue dots) with a Lorentzian lineshape fitting (red curve) augmented with mode splitting. The full-width-half-maximum (FWHM) linewidth is 210 MHz with splitting of 92 MHz corresponding to an intrinsic $\rm Q_0=1.7$ million and loaded $\rm Q_L=0.92$ M. Lower trace is a frequency calibration (yellow dots) from a fiber Mach-Zehnder Interferometer free spectral range is 191.26 MHz) with sinusoidal fitting (red curve).
{\bf(d)} Numerical simulation of normalized intracavity power versus normalized pump laser detuning ($2\delta\omega/\kappa=2(\omega-\omega_0)/\kappa$) for three AlGaAs microresonators having FSRs of 100 GHz, 300 GHz and 1 THz. The pump power is set to 36 times parametric oscillation threshold for all three traces. This corresponds to holding the normalized pumping parameter $f$ (see Methods) constant. The parameter $D_2/\kappa$ is set to the measured value of 1.1 for the 1 THz FSR simulation and scales quadratically with FSR in the other simulations (see Methods for details). }
\label{Fig1}
\end{figure*}

\medskip

\noindent\textbf{Microresonator Device Characterization}

The device was designed using an eigenmode solver with refractive index of Al$_{0.2}$Ga$_{0.8}$As taken from the material refractive index database \cite{Guden_1996}. Devices with 12.46 $\mu$m radius (FSR = 1 THz) and straight bus waveguides were fabricated (photomicrograph in Fig. \ref{Fig1}a). Details on the microresonator fabrication are provided in \citet{Xie2020AlGaAs}. A brief summary of the fabrication steps is also included in the Methods section.  The fundamental TE mode had a calculated effective index n$_{\textrm{eff}}$ of 2.899, group index n$_{\textrm{g}}$ of 3.746, and effective mode area A$_{\textrm{eff}}$ of 0.256 $\mu$m$^2$ at 1532 nm, the pumping wavelength in this study. 

A tunable ECDL (Toptica CTL1550) was used for both resonator characterization and for pumping of the soliton microcomb. The laser light is coupled in/out the chip bus waveguide through a pair of fiber lenses and pump power is adjusted by a variable optical attenuator (VOA) and monitored before and after the resonator chip by integrating sphere power meters (ISPMs). The ISPMs allowed estimate of the coupling loss to the chip to be 3.3 dB per facet. The bus waveguides were tapered to 200 nm to minimize this loss.

To characterize the mode family dispersion, mode frequencies were measured using the tunable laser with scan wavelength calibrated by an optical spectrum analyzer (OSA). Anomalous dispersion is verified in Fig. \ref{Fig1}b where parabolic fitting of the integrated dispersion (i.e., $\omega_\mu - \omega_0 - D_1 \mu \approx D_2 \mu^2 /2$) gives second-order dispersion $D_2/ 2 \pi$=235 MHz where $\mu$ is the mode number relative to pump mode frequency $\omega_0$, and $D_1 / 2 \pi$ = 1.013 THz is the FSR at the pump frequency. The measured microresonator properties correspond to negative group velocity dispersion (GVD) $\beta_{2}=-n_{g}D_2/cD_1^2$ $\approx -455$ ps$^2$/km and $D=-2\pi c\beta_{2}/\lambda^2 \approx$ 366 ps/nm/km, in good match with the prediction from modeling (346 ps/nm/km). As an aside, the combination of large dispersion and large FSR of this system leads to parametric oscillation initially on the neighboring sidemodes of the pumping mode as observed in microtoroids \cite{VahalaChi3OPO_PRL2004,TJK2007microcomb}. Specifically, the peak parametric gain occurs at $ \mu- \mu_{\textrm{pump}}\approx \sqrt{{\kappa}/{D_2}} \approx 1$ \cite{TJKDKSreview}.

\begin{figure*}[t!]
\centering
\includegraphics[width = \linewidth]{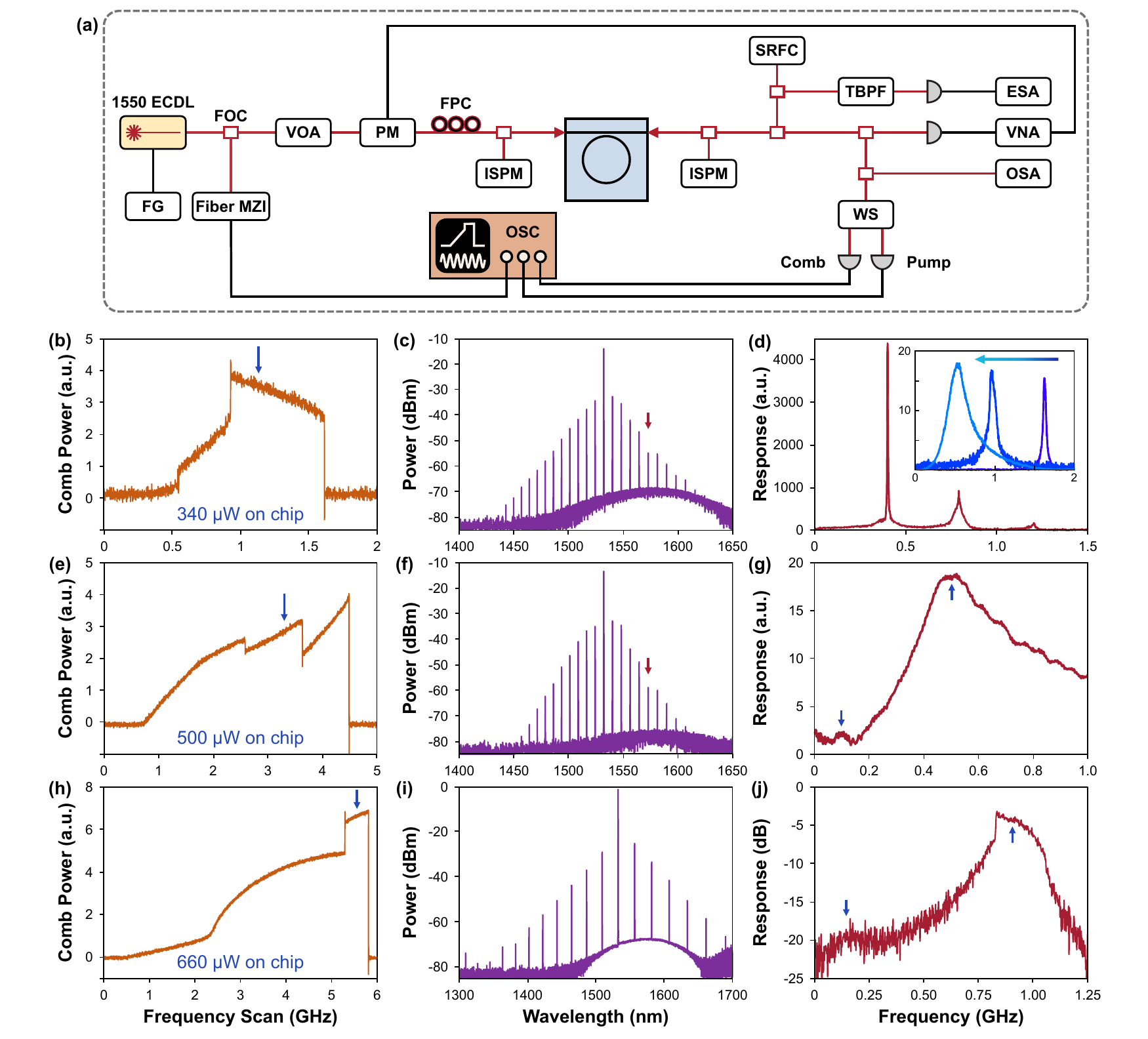}
\caption{\textbf{AlGaAs microresonator soliton generation behavior at different pump power levels. } {\bf (a)} Experimental setup. FG: Function Generator; ECDL: External Cavity Diode Laser;  FOC: Fiber Optic Coupler; Fiber MZI: Fiber Mach-Zehnder Interferometer; VOA: Variable Optical Attenuator; FPC: Fiber Polarization Controller; PM: Phase modulator; VNA: Vector Network Analyzer; WS: Wave Shaper; SRFC: Self-Referenced Fiber Comb; ESA: Electrical Spectral Analyzer; OSA: Optical Spectral Analyzer;  OSC: Oscilloscope; ISPM: InGaAs Integrating Sphere Photodiode Power Meter; PD: InGaAs Photodetector; TBPF: Tunable Band-pass Filter. (b)-(d) Breather soliton generation for 300 $\mu$W on-chip pump power showing soliton power  versus pump frequency scan (b), optical spectrum (c) at scan location given by the blue arrow in (b), and modulation response (d). (a.u. arbitrary units) Note that the $\mu=-5$ comb line (red arrow) is lower as a result of mode crossing (see Fig. 1b). In panel (d) breathing frequency peak and its harmonics are revealed. Inset: frequency response showing C-resonance tuning before tuning into soliton regime. The arrow shows the pump laser scan direction. (e)-(g) Single soliton formation at 500 $\mu$W on chip pump power with content similar to (b)-(d). The blue arrow indicates the single soliton state region and detuning location for spectrum measurement in panel f. In panel (g) the frequency modulation response of the single soliton state shows the S-resonance and C-resonance positions (blue arrows). 
(h)-(j) 3-FSR perfect soliton crystal formation at 660 $\mu$W on chip pump power. Panels are analagous to (e)-(g). }

\label{Fig2}
\end{figure*}

Intrinsic Q factor $\rm Q_0=1.7$ million and loaded Q factor $\rm Q_L=0.92$ M at the pumping wavelength 1532 nm were measured by characterizing the transmission spectrum as shown in Fig. \ref{Fig1}c. The transmission spectrum featured a full-width-half-maximum linewidth of $\kappa/2\pi=210$ MHz which was broadened slightly by back-scatter-induced splitting of 92 MHz  \cite{Kippenberg02OLbackscattering}. In the linewidth measurements, probe power was maintained low enough to avoid resonance thermal broadening  \cite{VahalaThermalLock_OE04}. The wavelength scan was also performed from blue to red wavelengths so that the measured Q values are conservative. 

The threshold for parametric oscillation can be estimated by the following equation \cite{VahalaChi3OPO_PRL2004,VahalaSilicaSoliton},
\begin{equation}
 P_{\textrm{th}}=\frac{\pi n_{\textrm{g}}\omega_0 A_{\textrm{eff}}}{4\eta n_2} \frac{1}{D_1Q^2}
\label{Threshold}    
\end{equation}
where $n_2$ is Kerr nonlinear coefficient, $\eta=Q/Q_{\textrm{ex}}$ is the waveguide-to-resonator loading factor, and other parameters are given above. Using $n_2\approx 1.7\times 10^{-17}$ m$^2$/W \cite{MDabsQ_NC22}, the parametric threshold is esimated to be $20 \;\mu$W, which agrees well with the measured in-waveguide threshold pump power of $22\;\mu$W.

\medskip

\noindent\textbf{Soliton Generation at Room Temperature}

Room temperature generation of solitons in AlGaAs is frustrated by a combination of large thermo-optic coefficient and the abrupt drop of intra-cavity power that usually accompanies soliton formation \cite{herr2014temporal}. This drop coincides with the transition of the system from the modulation instability regime to the soliton regime as a pump laser is wavelength tuned from blue to red into the soliton regime. The subsequent rapid cooling of the microresonator mode volume causes its frequency to quickly tune away from the pumping frequency. If the thermo-optic effect is very large, this tuning prevents stable soliton formation. On the other hand, the thermo-optic effect will also self-stabilize an optically pumped system if the intracavity power increases with laser blue-to-red tuning \cite{VahalaThermalLock_OE04}. For soliton generation, the stability benefit of this increase has been noted in \citet{PappCrystal_NP2017} and \citet{HydexSilicaPSC2018}. Specifically, when the soliton power is higher than power emitted in the modulation instability regime, the thermo-optic effect will assist (rather than oppose) stable soliton formation. As illustrated in Fig \ref{Fig1}d, this situation can happen when DKSs are operated at very high repetition-rates, which can be arranged by operating a resonator with many co-circulating solitons or by using a large FSR resonator in the single soliton regime.

\begin{figure}[t!]
\centering
\includegraphics[width = \linewidth]{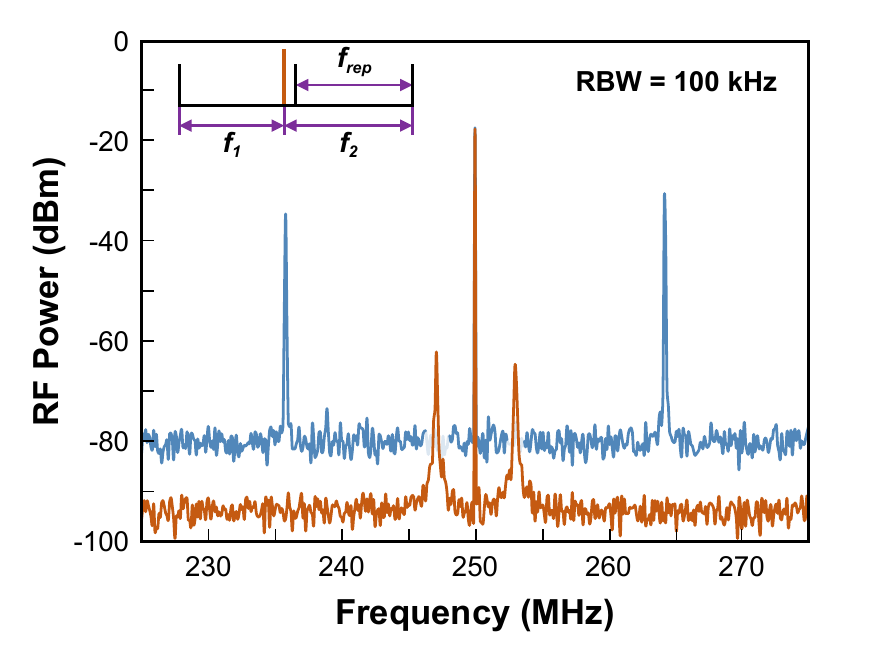}
\caption{\textbf{Single soliton comb line beatnote with self-referenced fiber comb.} The peak at 250 MHz gives the fiber comb repetition rate $f_{\textrm{rep}}$. In the red trace, the 2 small peaks are beatnotes between the filtered 1548 nm single soliton comb line with 2 fiber comb lines (see inset). The blue trace is the beatnote of the ECDL laser pump line with two fiber comb lines. }
\label{Fig3}
\end{figure}

To demonstrate this effect, the mode at 1532 nm was pumped for soliton generation. Details on the measurement setup are provided in Fig. \ref{Fig2}a.  A summary of comb behavior at three pump power levels is provided in Fig.\ref{Fig2} b-j. Beginning with panels Fig.\ref{Fig2} b-c, panel b gives the comb power for constant on-chip bus waveguide pump power of 340 $\mu$W as the pump laser is scanned in frequency (horizontal axis). A Mach-Zhender interferometer with a free-spectral-range of 191.26 MHz is used to calibrate the laser frequency scan. The pump line is filtered from the orange comb power trace using a WaveShaper.  The comb power shows an abrupt power jump where the system transitions from a modulation instability (MI) state to a comb state. As discussed earlier, because the comb power is here larger than the MI power, the photothermal effect works to stabilize the system. In fact, it is possible to manually tune the laser frequency into the comb state and stably measure comb spectra \cite{KartikTHz2017,TJKTHz2017}. A comb spectrum measured at the detuning indicated by the blue arrow in Fig.\ref{Fig2}b is presented in Fig.\ref{Fig2}c. The comb line spacing equals one FSR (1 THz, 8 nm). As an aside, the 1572 nm comb line (indicated by the red arrow) is reduced in power as a result of the perturbation to the parabolic dispersion at $\mu=-5$ in Fig. \ref{Fig1}b. 

To further characterize the comb, phase modulation (PM, Thorlabs LN65S-FC in Fig.\ref{Fig2}a) of the pump field was applied in combination with measurement of transmitted power using a vector network analyzer (VNA, Keysight E5061B in Fig.\ref{Fig2}a) \cite{TJKSwitching_NPhy17}.  Briefly, the PM is driven by the VNA with RF power of -7 dBm, and the drive frequency is scanned over the span of 1 MHz-3 GHz. Then, the transmitted light through the bus waveguide is detected by a fast photo detector (Thorlabs DXM30AF). The photodetector (PD) output signal is then input to the VNA for analysis. The measurement was repeated at each pumping power level and the resulting spectra are shown in Fig. \ref{Fig2}d,g,j.

The measured frequency response spectra in Fig. \ref{Fig2}d reveals this comb to correspond to a breather soliton  \cite{CYBreatherPRL2016,ErwanBreater_NC2017,MJBreather_NC2017}. Specifically, a sharp peak at breathing frequency 400 MHz  \cite{ErwanBreater_NC2017} is present with its 2nd and 3rd harmonic overtones. It is noted that modulation is not necessary for the peaks to appear, as the same peaks are observed in the base-band intensity noise spectrum. The soliton breathing frequency could be tuned from 290 MHz to 1 GHz by adjusting the pump-resonance detuning frequency.

By increasing pump power to 500 $\mu$W on chip, a single soliton state was measured as shown in Figs. \ref{Fig2}e-g. Manual tuning of the pump frequency was again used here. As can be seen in Fig. \ref{Fig2}e, the comb step initially decreases  slightly but eventually rises above the MI power so that once again the thermo-optic effect helps to stabilize the comb operating point. A comb spectrum is shown in Fig. \ref{Fig2}f corresponding to the frequency position given by the arrow in Fig. \ref{Fig2}e. The VNA measurement is shown in Fig. \ref{Fig2}g. The small spectral bump near 100 MHz is identified as the S-resonance \cite{TJKSwitching_NPhy17}. The C-resonance at 490 MHz is also apparent for this state.

In Figs. \ref{Fig2}h-j, the same measurements are repeated for 660 $\mu$W on chip pumping power. Here, the optical spectrum in Fig. \ref{Fig2}i reveals that the state is a perfect soliton crystal state  \cite{PappCrystal_NP2017,TJKPSC_NPhy19,HydexSynthesizedSC2021} with its repetition rate equal to 3 FSRs (3 THz, 24 nm) corresponding to three soliton pulses circulating in the cavity. The comb spectrum is spectrally broad spanning 400 nm (1300-1700 nm). Importantly, the ability to observe a soliton crystal further confirms the coherence of the soliton pulses in this system, since this is required to form the 3 FSR comb from coherent interference of three underlying 1 FSR combs. 

To further confirm microcomb formation in the single soliton state, one of the comb lines (1548 nm line in Fig. \ref{Fig2}f, $\mu=-2$) is filtered out and beat with a self-referenced  \cite{jones2000carrier} fiber comb (Menlo FC1500-250-ULN, $f_{\textrm{rep}}$=250 MHz and carrier-envelope-offset frequency $f_{\textrm{CEO}}$ = 35 MHz). The beatnote between microcomb and fiber comb is shown in Fig. \ref{Fig3} as the red trace. The strong peak is the fiber comb repetition rate, while the neighboring weaker peaks are beatnotes produced by two fiber comb teeth with the filtered THz comb tooth. The beatnote peak at 247.1 MHz has a full width at half maximum linewidth of 59 kHz, which is an order-of-magnitude narrower than in previous work at 4K temperature  \cite{KartikAlGaAsLTDKS_LPR20}. For comparison, the beatnote between the pump laser (Toptica ECDL) and the fiber comb is also plotted as the blue trace. The microcomb beatnote is visibly wider than the pump laser beatnote. The specific source of noise in this system is not known, but could be related to pump frequency transduction noise which is known to impact comb repetition rate in other systems \cite{XuYiquietpoint,QifanDWnoise}. Such repetition rate noise is multiplied with increasing comb tooth number away from the pumping line \cite{lei2022optical}.

As an aside, on account of the THz repetition rate, it was difficult to directly detect the comb rate \cite{PascalDHayeToriodOE19}. Likewise, autocorrelation measurements were difficult due to the sub femto Joule pulse energy. Also, amplification of the pulses was challenging due to the limited bandwidth of erbium fiber amplifiers in comparison to the soliton spectrum.

\medskip

\noindent\textbf{Summary and Discussion}

We have demonstrated soliton generation in AlGaAs microresonators at room temperature for the first time. In the experiment, we used large FSR resonators to create rising soliton steps and leverage the previously-reported large thermo-optical effect. Low-noise 1 THz repetition-rate soliton generation was possible with only sub-milliWatt optical pump power. Combined with on-chip optical amplifiers\cite{liu2022photonic},  such a high repetition-rate optical pulse source can be useful in applications such as pulse driven optical computation \cite{markov2014limits,feldmann2019all} (e.g. coherent Ising machine  \cite{yamamoto2020coherent,mcmahon2016fully,100kspinCIM}) and for THz wave generation \cite{PascalDHayeToriodOE19}. It should also be possible to integrate these devices with III-V pump lasers.

\noindent

\noindent
\medskip
\begin{footnotesize}
%\begin{methods}

% \noindent \textbf{Funding Information}: 

% \noindent\textcolor{blue}{This work was supported by NTTRI.}

\noindent \textbf{Acknowledgments}: 

\noindent %The authors  thank NTT Research for their financial and technical support. 
The authors acknowledge  Chengying Bao (THU), Maxim Karpov (Enlightra) and Qifan Yang (PKU) for fruitful discussions. A portion of this work was performed in the UCSB Nanofabrication Facility, an open access laboratory.

\vspace{3mm}

\noindent \textbf{Author Information}: 

\noindent Weiqiang Xie 's current address is Department of Electronic Engineering, Shanghai Jiao Tong University, 800 Dongchuan Road, Shanghai, 200240, China.

\end{footnotesize}

\bibliography{bibliography}

\clearpage

\begin{large}
\noindent\textbf{Methods} 
\end{large}

\medskip

\noindent\textbf{Device fabrication}

 The Al$_{0.2}$Ga$_{0.8}$As was grown by molecular beam epitaxial (MBE) on a 3 inch GaAs substrate, which was then bonded to a 3 $\mu$m thick thermal oxide layer grown on a 4 inch Si substrate by thermal oxidation. The GaAs substrate was subsequently removed by the etch-stop layer technique using a 500 nm Al$_{0.8}$Ga$_{0.2}$As layer. A 100 nm thick SiO$_{2}$ layer was deposited using atomic layer deposition (ALD) as a dry etch hard mask. The microresonator pattern was generated by deep-ultraviolet (DUV) photolithography (248 nm). The photoresist after development was then reflowed at 155$^\circ$C to reduce pattern line edge roughness (LER)  \cite{Xie2020AlGaAs}.  The pattern was transferred to the SiO$_{2}$ hard mask by inductively coupled plasma reactive-ion etching (ICP-RIE) with a mixture of CHF$_{3}$/CF$_{4}$/O$_{2}$. The Al$_{0.2}$Ga$_{0.8}$As was then etched through by ICP-RIE with a mixture of Cl$_{2}$/N$_{2}$ having high etching selectivity ($>$10 selectivity of AlGaAs over SiO$_2$) to make the vertical sidewall. Surface passivation of the AlGaAs waveguides was implemented by depositing a 5-10 nm Al$_{2}$O$_{3}$ layer and a 50-nm SiO$_{2}$ layer using ALD. The passivation  reduces absorption caused by surface states. The wafer was then cladded with 1.5 $\mu$m thick SiO$_{2}$ using  plasma-enhanced chemical vapor deposition (PECVD) and diced into device chips with dicing saw line aligned with the 200 nm waveguide tapering region to enable low facet coupling loss. 

\medskip

\vspace{2mm}
\noindent\textbf{Numerical Simulations}

 The slowly varying (slower than the round trip) intracavity field $A(\phi,t)$ in the microresonator can be described by Lugiato-Lefever equation (LLE) \cite{LLE}:
\begin{equation}
   \frac{\partial A}{\partial t}=-(\frac{\kappa}{2}+i\delta \omega)A +i\frac{D_2}{2}\frac{\partial^2 A}{\partial \phi^2}+ig|A|^2 A+\sqrt{\frac{\kappa\eta P_{\textrm{in}}}{\hbar \omega_0}}
\end{equation}
where  $g=\hbar \omega^2_0 n_2 D_1/2\pi n_g A_{\textrm{eff}} $ denotes the normalized Kerr effect nonlinear coefficient, $t$ is the slow time , $\phi$ is the angular coordinate in the ring resonator in a frame copropagating with the soliton, $\kappa$ is the cavity total loss rate, $\delta\omega=\omega-\omega_0$ is the pump cavity detuning ($\omega$ is the pump laser frequency and $\omega_0$ is the cavity resonance frequency), $D_2$ is the cavity group velocity dispersion (GVD), $\eta=\kappa_{\textrm{ex}}/\kappa$ is the cavity coupling factor (i.e., waveguide coupling rate $\kappa_{\textrm{ex}}$ normalized to total loss rate $\kappa$), and $\hbar$ is the reduced Planck constant. The thermo-optical effect and mode splitting are not included in this equation. 

To generate Fig. \ref{Fig1}b in the main text, the resonator diameter is varied to obtain resonator FSR (D$_1$/2$\pi$) values of 100 GHz, 300 GHz and 1 THz. The resonator intrinsic loss ($\kappa_{\textrm{in}}$) and loading ($\eta$) are fixed so that the total loss ($\kappa$) is fixed. The resonator transverse mode properties (A$_{\textrm{eff}}$, n$_{\textrm{g}}$ and $\beta_2$) are also fixed.  However, since the group velocity dispersion $\beta_2=-n_gD_2/cD_1^2$ (GVD) is constant, $D_2$ scales quadratically with $D_1$. 
Also, the parametric oscillation threshold $P_\textrm{th}$ scales inversely with $D_1$. Therefore, in plotting Fig. \ref{Fig1}b, we have set the input pump power P$_\textrm{in}$ to scale inversely with D$_1$.  This has the effect of holding the normalized pumping parameter $f$ (see below) constant as $D_1$ is varied. It also maintains a constant soliton step length in the plots since the soliton existence range depends upon $f$. 

For numerical simulation, it is  convenient to normalize the LLE equation \cite{TJKDKSreview,yixuthesis} by taking $\tau=t\cdot\kappa/2$ and $\Psi=\sqrt{2g/\kappa}A$, where $\tau$ is normalized time,  and $\Psi(\phi,\tau)$ is the normalized slowing varying optical field, the following equation results,

\begin{equation}
   \frac{\partial \Psi}{\partial \tau}=-(1+i\zeta)\Psi +i\frac{D_2}{\kappa}\frac{\partial^2 \Psi}{\partial \phi^2}+i|\Psi|^2 \Psi+f
\end{equation}
where $\zeta=2\delta\omega/\kappa$ is the normalized cavity detuing and $ f=\sqrt{8\eta g P_{\textrm{in}}/\kappa^2\hbar \omega_0}$ is the normalized pumping term that, as noted above, is held constant as $D_1$ is varied. Also, $\phi$ remains the angular coordinate after normalization. Since $f$ is held constant, only $D_2/\kappa$ in Eq.3 determines the system behavior. Again, $D_2$ scales quadratically with $D_1$ for constant GVD and produces different behaviors in simulation presented in Fig. \ref{Fig1}b. 

\vspace{2mm}

\end{document}